\documentclass[12pt,german]{article}
\begin{document}

\sloppy
\newcommand{\eqb}{\begin{equation}}
\newcommand{\eqe}{\end{equation}}
\newcommand{\dmb}{\begin{displaymath}}
\newcommand{\dme}{\end{displaymath}}
\newcommand{\pd}{\partial}
\newcommand{\eab}{\begin{eqnarray}}
\newcommand{\eae}{\end{eqnarray}}
\newcommand{\eabw}{\begin{eqnarray*}}
\newcommand{\eaew}{\end{eqnarray*}}
\newcommand{\la}{\langle}
\newcommand{\ra}{\rangle}
\newcommand{\ep}{\varepsilon}
\pagestyle{myheadings}

\begin{center}
\Large
{\bf Direct Test of the Scalar-Vector Lorentz Structure of the Nucleon- and Antinucleon-Nucleus Potential}
\normalsize
\end{center}
\begin{center}
R. Hofmann\footnote{Supported by the Graduiertenkolleg under contract DFG GRK 132/3}, 
Amand Faessler\footnote{Supported by BMBF 06 T\"u 887} and 
Th. Gutsche$^2$
\\[12pt] 
{\it Institut f\"ur Theoretische Physik, Universit\"at T\"{u}bingen\\
Auf der Morgenstelle 14, D--72076 T\"{u}bingen, Germany}\\
\end{center}
\begin{center}
{\bf Abstract}
\end{center}
\begin{small}
Quantum Hadrodynamics in mean field approximation describes the effective 
nucleon-nucleus potential (about -50 MeV deep) 
as resulting from a strong repulsive vector (about 400 MeV) and 
a strong attractive scalar (about -450 MeV) contribution. This scalar-vector Lorentz structure implies a 
significant lowering of the threshold for $p\bar{p}$ photoproduction on a nucleus by about 850 MeV 
as compared to the free case since charge conjugation reverses the sign of the vector potential contribution 
in the equation of motion for the $\bar{p}$ states. It also implies a certain size of the photon induced 
$p\bar{p}$ pair creation cross section near threshold which is calculated for a target nucleus $^{208}$Pb. 
We also indicate a measurable second signature of the $p\bar{p}$ photoproduction process by
estimating the increased cross section for emission of charged pions as a 
consequence of $\bar{p}$ annihilation within the nucleus.


\end{small}
\vspace{.8cm}  

\section{Introduction}

In this article we propose a direct method to investigate if Quantum Hadrodynamics (QHD I,II)\ \cite{Wa1},  
 which is well suited for the description of nuclear properties such as density distributions, 
binding energies etc., 
can be extrapolated to the explicit consideration of antinucleons inside heavy nuclei. In its 
simplest version, that is QHD I, the
relativistic nuclear many-body problem is based on a Lagrangian containing nucleon and isoscalar meson 
(scalar meson $\sigma$ and vector meson $\omega$) degrees of freedom. 
In the mean field Hartree approximation \cite{Wa1} nucleon single particle states are determined from 
a Dirac equation with Lorentz scalar and vector mean field potentials resulting in positive and 
negative frequency solutions. According to hole theory  
the physical vacuum is characterized by a completely filled negative energy Dirac sea 
and the absence of states carrying positive energy. The charge conjugation 
operation allows for the interpretation of holes in the sea as antiparticles. 

While the scalar mean field is attractive for both nucleon and antinucleon states, the potential induced by the vector meson 
reverses sign under charge conjugation. Hence  
scalar and vector potentials for the nucleon almost cancel, resulting in an effective potential 
depth of about -50 MeV, while both contributions add up for the antinucleon 
generating an effective potential depth of -700 to -900 MeV. 
 
In this article we investigate the consequences of the Lorentz structure of the 
nucleon/antinucleon ($N\bar{N}$) mean field potentials,
as set up in QHD I, on predictions for near threshold cross sections of 
$\gamma$ induced proton-antiproton ($p\bar{p}$) pair creation. 
The respective threshold is determined by the energy of the lowest lying $\bar{p}$ bound state and the Fermi energy 
of the nuclear ground state. As compared to
 the free case the $p\bar{p}$ pair creation threshold is therefore strongly reduced due to 
the large attractive potential felt by the $\bar{p}$ (for a schematic description see Fig. 1). 
At present it is not clear whether the mean field approximation of QHD I delivers a sufficient 
description of the $\bar{N}$-nucleus potential since $N\bar{N}$ annihilation processes 
can generate a dispersive complex contribution to the mean field potential.  

By studying the photoproduction of $p\bar{p}$ pairs on a nucleus we directly test
 the highly contested assumption of a mean field for the description 
 of explicit antinucleons inside the nucleus. 
 The question if this assumption is reasonable or if one is 
forced to include effects beyond the mean field approach can in principle be answered by experimental tests of the
$p\bar{p}$ pair creation process on nuclei.  

For the $\gamma$-induced $p\bar{p}$ pair creation we consider a process where a low energy proton 
is emitted into the continuum while in a first step the $\bar{p}$ remains in a bound state in the nucleus.
 There are, of course, a variety of competing reactions that also lead to final state protons. 
At a $\gamma$-energy of several hundred MeV (hence well below the $p\bar{p}$ pair creation threshold) 
Intra Nuclear Cascade (INC) calculations for $\gamma$ absorption on nuclei \cite{Oset1,Oset2} are able to reproduce 
the experimental cross sections. For photon energies of about 1000 MeV, lying around the $p\bar{p}$ pair creation 
threshold, there exists a semiclassical BUU transport model calculation \cite{Eff} which  
gives an order of magnitude estimate for the 
background due to the direct photoemission of protons. 

The $\bar{p}$, produced in a bound state, can subsequently annihilate in the nucleus, thereby producing 
a definite average number of emitted pions \cite{Raf}. 
For the emission of charged pions this number has been calculated in the framework of an INC calculation 
by Botvina et al. \cite{Bot} and turned out to be 2.5 for heavy nuclei. 
Given the calculated $\gamma$-induced $p\bar{p}$ pair 
creation cross section on nuclei both the cross section of proton and pion emission  
due to the annihilation of the $\bar{p}$ can be estimated. 
These results are compared to the respective theoretical estimates for the background of
protons and pions coming from direct $\gamma$-induced emission. 
   
The paper is organized as follows: 
In the next section we will shortly describe the derivation of the static mean field potentials from the 
Lagrangian of QHD I\ \cite{Wa1} and discuss the relevant parts of the nucleon and antinucleon spectrum. 
Here we consider only the simplest form of QHD that allows for interactions of the nucleon with 
a neutral scalar and vector meson field while neglecting the electromagnetic and additional vector meson mediated 
interactions as described in Ref.\cite{Wa2}. 
In Section 3 we indicate the construction of the S-matrix in first order perturbation theory 
for the $\gamma$-induced creation of a $p\bar{p}$ pair on nuclei, with details of 
the actual calculations contained in the Appendices. 
The dependence of the $\gamma$-induced $p\bar{p}$ pair creation cross section on the 
depth of the scalar-vector potential 
for a given set of quantum numbers and the results for differential and total inclusive cross sections 
are presented in Section 4. Finally, in Section 5, we summarize the results and give the conclusions.

\section{Mean field potentials}

The Lagrangian of QHD I \ \cite{Wa1} (also refered to as Walecka Model) 
for interacting nucleon ($\psi$), massive scalar meson ($\phi$) and 
vector meson ($V_{\mu}$) fields is given by
\eab
\label{waL}
{\cal L}_{WM}&=&\bar{\psi}[\gamma^\mu(i\pd_\mu-g_vV_\mu)-(M-g_s\phi)]\psi
+\frac{1}{2}(\pd^\mu\phi\pd_\mu\phi-m_s^2\phi^2)\nonumber\\  
& &-\frac{1}{4}F^{\mu\nu}F_{\mu\nu}+\frac{1}{2}m_v^2V^{\mu}V_{\mu}\ \ ,
\eae
where the field strength tensor for the vector meson is defined by
\eqb
F^{\mu\nu}=\pd^{\mu}V^\nu-\pd^{\nu}V^\mu\ \ .
\eqe
According to Ref.\cite{Wa1}, in a nuclear system with a large number of 
nucleons at nuclear saturation density one 
can consider the meson fields as classical fields $V^c_\mu, \phi^c$ 
and only quantize the nucleon field. 
For spherically symmetric, static nuclei the equations of motion for the meson fields are
\eab
\label{mesonem}
(\bigtriangledown^2-m_s^2)\phi^c(r)&=&-g_s\rho_S(r)\nonumber\\ 
(\bigtriangledown^2-m_v^2)V^c_0(r)&=&-g_vj_0(r)
\eae
while the spatial part of $V^c_\mu$ must vanish due to rotational symmetry.
The scalar and current densities of Eq.(\ref{mesonem}) are the normal ordered expectation values of the 
corresponding field operators taken in 
the nuclear Hartree ground state $|F\ra$
\eqb
\rho_S(r):=\la F|:\bar{\psi}\psi:|F\ra,\ \ 
 j_\mu(r):=\la F|:\bar{\psi}\gamma_\mu\psi:|F\ra\ \ .
\eqe
The Dirac equation for the baryon field modes reads 
\eqb
\label{barem}
[i\gamma^\mu\pd_\mu-g_v\gamma^0V^c_0(r)-(M-g_s\phi^c(r))]\psi=0\ .
\eqe
Solutions to the coupled system of Eqs.(\ref{mesonem}) and (\ref{barem}) can be found by the usual 
Hartree procedure, where experimental information about nuclear bulk properties is used 
to determine the coupling constants 
$g_v$ and $g_s$ \cite{Wa1}.
 
In the following we will assume the potential shapes for $\phi^c(r)$ and $V^c_0(r)$ 
to be approximated by spherical square wells as
\eab
\label{S}
\phi^c(r)&=&\left\{
\begin{array}{ll}
S\ \ ,\ \ \ \ \ \ \ \ \ \ \ \ \ \ \ \ \ \ \ \ 0\le r\le r_0&\\
0\ \ ,\ \ \ \ \ \ \ \ \ \ \ \ \ \ \ \ \ \ \ \ \ r_0<r\  , &
\quad 
\end{array}
\right.\nonumber\\ 
V^c_0(r)&=&\left\{
\begin{array}{ll}
V\ \ ,\ \ \ \ \ \ \ \ \ \ \ \ \ \ \ \ \ \ \ \ 0\le r\le r_0&\\
0\ \ ,\ \ \ \ \ \ \ \ \ \ \ \ \ \ \ \ \ \ \ \ \ r_0<r\  . &
\quad 
\end{array}
\right.
\eae
Although this is a rough approximation to the more realistic potentials obtained in Ref.\cite{Wa1} it bares the
 advantage that analytical solutions are obtained for Eq.(\ref{barem}). 
 For a first estimate of the $\gamma$-induced $p\bar{p}$ pair creation cross section 
 the approximation of Eq.(\ref{S}) is sufficient.
 
The eigenvalues of Eq.(\ref{barem}) are determined by demanding continuity for 
the single particle solutions $\psi$ at $r=r_0$. For the nucleus $^{208}$Pb we choose potential 
strengths of $S=400$ MeV, $V=450$ MeV and a 
nuclear radius given by 
\dmb
r_0\approx1.2\times A^{\frac{1}{3}}\ \mbox{fm=7.2\ fm}\ . 
\dme
The corresponding spectrum of the negative energy solutions of Eq.(\ref{barem}) is indicated in Fig. 2, 
wherethe $\kappa$ denotes the Dirac quantum number. The potential parameter values are compatible with those determined 
by a Thomas-Fermi approximation for finite systems and a relativistic Hartree calculation as outlined 
in Refs.\cite{Wa1, Wa2}.

\section{S-matrix element for $\gamma$-induced $p\bar{p}$ pair creation}

In the following we indicate the calculation of the S-matrix element for $\gamma$-induced $p\bar{p}$ 
pair creation to first order in the electromagnetic coupling constant $e$. 
Here we assume the proton
to be in a continuum state and the antiproton to be produced in a bound state (see Fig. 1). 
For simplicity we assume that the proton state is approximated by a positive energy plane wave 
\eqb
\label{pro}
^{(+)}\psi^\mu_p(x)=\sqrt{\frac{M}{E({\vec p})V}}u({\vec p},\mu)e^{-ipx},
\ \ \ \ E(\vec p):=p_0=\sqrt{\vec p^{\ 2}+M^2}\ .
\eqe
The $\bar{p}$ state is associated with a negative energy solution of Eq.(\ref{barem}) and given by 
\eqb
\label{apro}
^{(-)}\psi^\rho_{p^\prime}(x)=\exp(-iE^\prime x^0)\left(
\begin{array}{ll}
g(r,E^\prime)\ \omega^\rho_\kappa(\theta,\phi)\\
if(r,E^\prime)\ \omega^\rho_{-\kappa}(\theta,\phi)
\end{array}
\right)\ ,\ \ \ \ E^\prime<0\ .
\eqe
Here $\rho$ denotes the total angular momentum projection, $\mu$ the 
spin projection, $p^\prime$ summarizes energy and Dirac 
quantum number $\kappa$, while $p$ denotes the respective four momentum. 
 With Bjorken-Drell convention for the $u({\vec p},\mu)$ spinor, the wave function
 in Eq.(\ref{pro}) 
is normalized to unit probability within a volume $V$. Accordingly, 
the same normalization is applied to the wave function of Eq.(\ref{apro}), with details given in Appendix A. 
The S-matrix element for $\gamma$-induced $p\bar{p}$ pair creation in first order perturbation theory is then 
deduced as
\eab
\label{smat}
S^{1,2;\mu;\rho}_{p,p^\prime}&=&ie\int d^4x\  ^{(+)}\bar{\psi}^\mu_p(x) 
A_\nu^{1,2}(x){\gamma^{\nu}}\ ^{(-)}\psi^\rho_{p^\prime}(x)\nonumber\\ 
&=&ie\int d^4x\ A_\nu^{1,2}(x)\ j^\nu_{tr}\ .
\eae
The potential $A_\nu^{1,2}(x)$ for a free photon propagating in 3-direction 
can  be written in Lorentz gauge in the following form
\eqb
\label{Anu}
A_\nu^{1,2}(x)=\frac{\ep_\nu^{1,2}}{\sqrt{2\ kV}}\ e^{-ik^\prime x}, 
\ \ \ \  k^{\prime \ 0}=|\vec k^\prime|=:k,\ \ep_\nu^1=(0,1,0,0), \ \ep_\nu^2=(0,0,1,0)\ ,
\eqe
where the four vectors $\ep_\nu^{1,2}$ characterize 
the two independent transversal polarizations. The potential $A_\nu^{1,2}$ is 
normalized to energy $k$ for the volume V. 

In contrast to the $\gamma$-induced $p\bar{p}$ creation in free space 
the S-matrix element of Eq.(\ref{smat}) does not vanish since the
negative energy eigenstate in Eq.(\ref{apro}) has a nontrivial expansion with respect to momentum eigenstates. 

Since the created $p$ and $\bar{p}$ carry inner structure the transition current $j^\nu_{tr}$ of 
Eq.(\ref{smat}) has to be modified. 
By including additional 
vector meson interactions in the Lagrangian of Eq.(\ref{waL}) the electromagnetic form 
factor of the nucleon can be simulated 
via vector meson dominance \cite{Sak, Wa2}. We disregard this possibility 
by using the simple $\sigma$-$\omega$-model of QHD I.  A minimal correction to 
the transition current $j^\nu_{tr}$ of Eq.(\ref{smat}) that should not be neglected
is the contribution of the anomalous magnetic moment $\mu_{a.m.}$ of the proton. 
For an asymptotically free theory the demand for Lorentz covariance (parity transformations included) 
and current conservation leads to the following form of the corrected transition current \cite{Itz}
\eqb
\label{trc} 
j^\nu_{tr,\ cor}=\bar{\psi}_q\gamma^{\nu}
\psi_{q^\prime}+\frac{\mu_{a.m.}}{2M}\pd_\lambda\left(\bar{\psi}_q
\sigma^{\nu\lambda}\psi_{q^\prime}\right)\ ,
\eqe
where $q$ and $q^\prime$ denote the four momenta of the corresponding states. The constant $\mu_{a.m.}$ is 
determined in the nonrelativistic limit as 
\eqb
\mu_{a.m.}=1.79284\ .
\eqe
In Appendix B it is shown that Eq.(\ref{trc}) even
holds for eigenstates of the Dirac equation with scalar-vector potentials
 if one demands some plausible properties for 
$j^\nu_{tr,\ cor}$. By partial integration we identify
\eqb
\frac{\mu_{a.m.}}{2M}\int d^4x\  A_\nu\pd_\lambda\left(\bar{\psi}_p \sigma^{\nu\lambda}\psi_{p^\prime}\right)
=-\frac{\mu_{a.m.}}{2M}\int d^4x\ \left(\pd_\lambda A_\nu\right)
\bar{\psi}_p\sigma^{\nu\lambda}\psi_{p^\prime}\ ,
\eqe
which leads to the corrected S-matrix element including the contribution of the anomalous magnetic moment as
\eab
\label{smatkorr}
^{cor}S^{1,2;\mu;\rho}_{p,p^\prime}&=&ie\int d^4x\  ^{(+)}\bar{\psi}^\mu_p(x) 
A_\nu^{1,2}(x)\left[{\gamma^{\nu}}+i\frac{\mu_{a.m.}}{2M}k_\lambda\sigma^{\nu\lambda}
\right]\ ^{(-)}\psi^\rho_{p^\prime}(x)\nonumber\\ 
&=&ie\int d^4x\ A_\nu^{1,2}(x)\ j^\nu_{tr,\ cor}\ .
\eae
The subsequent standard derivation of the differential cross section $\frac{d\sigma}{d\Omega}$
 for $\gamma$-induced $p\bar{p}$ pair creation can be found 
in Appendix C.

\section{Results}

The sensitivity of the $\gamma$-induced $p\bar{p}$ pair creation cross section $\sigma$ 
on the depths of the scalar and vector potentials is displayed in Fig. 3. 
Here the cross section $\sigma$ is given for a final state proton at fixed
 energy (M+8 MeV, M is the nucleon mass), while the $\bar{p}$ corresponds to the highest lying negative energy 
 state with $\kappa=+1$. In the variation of the potential depths the difference $\Delta:=S-V=50$ MeV is 
kept fixed and the value of $r_0$ has been chosen as $r_0=7.2$ fm. The two maxima of $\sigma$ in Fig. 3 
correspond to $S=380$ MeV and $S=460$ MeV and 
reflect the maximal overlap of the bound negative energy solution (associated with the 
lowest bound $\bar{p}$ state) with the continuum wave function for the $p$ state. 

The following results for photoproduction of $p\bar{p}$ have been obtained 
using the mean field potential parameter 
values introduced in section 2, which is an adaption 
of the $^{208}$Pb nuclear system (compare to Ref.\cite{Wa2}). In Fig. 4 we indicate the 
differential inclusive cross sections $\frac{d\sigma_{inc}}{d\Omega}$ for $p\bar{p}$ pair 
creation at $\gamma$-energies $k$ of 50 MeV and 30 MeV above threshold energy $k_{th}$. 
One obtains $\frac{d\sigma_{inc}}{d\Omega}$ by summing up the contributions of all energetically possible states 
of the produced $p\bar{p}$ pair. The global maximum of 
$\frac{d\sigma_{inc}}{d\Omega}$ for $k=k_{th}+30$ MeV is at $\theta=0$, while for $k=k_{th}+50$ MeV
 we obtain a pronounced global maximum at $\theta\approx\frac{\pi}{4}$. 

The integrated inclusive cross section $\sigma_{inc}$ for 
 direct proton emission due to $p\bar{p}$ pair creation as a function of photon energy $k$ 
 in the range from $k_{th}$ to $k_{th}+70$ MeV 
is shown in Fig. 5. For this energy range $\bar{p}$ bound states with 
Dirac quantum numbers up to $\kappa=+7$ are included. Final state protons with maximal kinetic energy 
give the largest contribution to $\sigma_{inc}$ due to favourable phase space (see for example Eq.(\ref{cs})).
 At a $\gamma$ energy $k$ of about 70 MeV above threshold we obtain an inclusive cross 
 section of $10^{-3}$ mb to $10^{-2}$ mb for proton emission due 
 to $p\bar{p}$ pair creation. 
 The background $\sigma_{bg}^p$ of conventional $\gamma$-induced proton emission is 
 estimated at 50 mb from a BUU transport 
 model calculation at a photon energy of $k=1$ GeV \cite{Eff}. 

Assuming that in each channel of the annihilation process 
\dmb
\bar{p}+N\longrightarrow X
\dme
there is at least a charged pion in the final state we have 
for the total inclusive pion emission cross section $\sigma_{inc}^{ch. pion}$ at $k=k_{th}+70$ MeV 
\dmb
\sigma_{tot,inc}^{ch. pion}(k_{th}+70\ \mbox{MeV})
\approx\sigma_{inc}(k_{th}+70\ \mbox{MeV})=10^{-3}\ \mbox{to}\ 10^{-2}\ \mbox{mb}\ ,
\dme
where $\sigma_{ tot,inc}^{ch. pion}$ is obtained by summing the inclusive cross sections 
$\sigma_{inc}^{ch. pion}$ for specific channels over all channels. Beside the increasing pion 
production total inclusive cross section due to $N\bar{p}$ annihilation 
in the $^{208}$ Pb nucleus an additional observable  
is the average multiplicity of 2.5 emitted charged pions \cite{Bot}, which 
can in principle be experimentally tested by coincidence measurements. 
The corresponding background $\sigma_{bg}^{ch.pion}$ of conventional 
charged pion emission is estimated at 15 mb 
 based on a BUU transport model calculation at a $\gamma$-energy of $k=1$ GeV  \cite{Eff}.

\section{Conclusions}

Quantum Hadrodynamics I \cite{Wa1, Wa2} solved in mean field Hartree approximation 
yields a scalar-vector structure for the real $N$-nucleus potential, where the 
effective potential depth ($\approx -50$ MeV) results from the difference of 
a large repulsive vector ($\approx 400$ MeV) and a large attractive scalar ($\approx 450$ MeV) potential. 
If one takes the simple mean field picture, according to hole theory, the $\bar{N}$-nucleus 
potential would effectively be about 
$850$ MeV deep. This approach neglects typical quantum effects of the full theory and by construction any 
correlations between the nucleons in the nuclear
ground state. Already on the level of a two particle problem 
there are essential contributions to the imaginary and dispersive real part of the antinucleon-nucleon  
potential due to $N\bar{N}$-annihilation. Therefore the mean field picture is highly contested. 

For example, Teis et al. \cite{Teis}
 indicate the lack of unitarity between the real and
imaginary part of the $\bar{N}$ self energy and the absence of Fock terms 
when applying the simple charge conjugation picture to construct the $\bar{N}$-nucleus potential in the simple  
Hartree meson mean field approximation. On the contrary, Mishustin et al. \cite{Mis} and Schaffner et al. \cite{Scha} 
argue for spontaneous and induced creation of 
$N\bar{N}$ pairs subject to scalar-vector nuclear mean fields in heavy ion collisions. We therefore propose a direct test 
of the Hartree mean field picture, that is of the depths of the scalar and vector $\bar{N}$-nucleus potentials. 
This can be done by measuring the near threshold photoproduction of $p\bar{p}$ pairs on a nucleus, here specified 
for $^{208}$ Pb. 

In this work we calculated the inclusive 
cross section for $\gamma$-induced $p\bar{p}$ pair creation at $\gamma$-energies up to about 
70 Mev above threshold. Thereby we assume that the 
Hartree mean field potential is approximated by a square well (adapted to $^{208}$Pb), while 
for the proton continuum states we use undistorted plane waves. The inclusive cross 
section is of the order of $10^{-3}$ to  $10^{-2}$ mb for $\gamma$-energies of 70 MeV above threshold. 

Transport model (BUU) calculations at a photon 
energy of about 1 GeV  \cite{Eff} for proton emission excluding  $p\bar{p}$ 
pair creation processes estimate the background cross section $\sigma_{bg}^p$ at about 50 mb 
for the $^{208}$Pb nuclear system. 
Given $\sigma_{bg}^p$, a $4\ \sigma$ error confidence interval 
and the calculated inclusive cross section of 
$\sigma_{inc}(70$MeV+$k_{th})\approx10^{-2}$  mb would at least require 
$4\times 10^8$ events $\mu_p$ of the inclusive
reaction
\dmb
\gamma+^{208}\mbox{Pb}\longrightarrow p+X
\dme
to isolate the $p\bar{p}$ 
creation signal from statistical fluctuations
\eab
\frac{4\ \sqrt{\mu_p}}{\mu_p}&\le&\frac{\sigma_{inc}(70\ \mbox{MeV}+k_{th})}{\sigma_{bg}^p}
\ \Longrightarrow\nonumber\\ 
\mu_p&\ge&\left (\frac{4\ \sigma_{bg}^p}{\sigma_{inc}(70\ \mbox{MeV}+k_{th})}\right)^2
\nonumber\\ 
&=&4\times 10^8\ .
\eae
This amounts to a measuring time of about 30 days if one takes a photon flux per bin (10 MeV) of 
$10^5\ \mbox{s}^{-1}$ and 
a $^{208}$Pb target with area mass density of 11g/cm$^2$. 
The photon flux is compatible with values obtainable at CEBAF \cite{Grab}.  

 A second signature for the $\gamma$-induced $p\bar{p}$ pair production is the 
emission of charged pions as produced by the strong annihilation of the 
$\bar{p}$ with a nucleon of the nucleus. For $^{208}$Pb the background cross section 
$\sigma_{bg}^{ch.pion}$ of the reaction
\dmb
\gamma+^{208}\mbox{Pb}\longrightarrow \pi^{\pm}+X
\dme
is estimated with the help of a 
transport model (BUU) calculation \cite{Eff} to be 15 mb at a $\gamma$ energy of 1 GeV. This would require 
an event number of measured charged pions greater than $3.6\times 10^7$ to suppress statistical fluctuations
sufficiently. 
With the above photon flux per bin and target area density the measuring time is estimated at about 9 days. The
predictions of the transport model calculation for the average multiplicity of charged pions due to conventionell 
photon absorption is about 0.35 \cite{Eff}. On the other hand, an INC calculations for the 
multiplicity of charged pions due $\bar{p}$ annihilation in the $^{208}$Pb nucleus suggests a value of 2.5 \cite{Bot}.
Thus in a coincidence experiment one would expect to detect a shift of the multiplicity towards higher values when
crossing the $p\bar{p}$ pair creation threshold from below. This effect is small since the ratio 
\dmb
\frac{\sigma_{inc}^{ch. pion}}{\sigma_{bg}^{ch. pion}}
\dme
is small and one again would need event numbers of the above order to isolate 
the signal from statistical fluctuations. 

To summarize, the observation of the emission of charged pions (cross section and multiplicity) 
  seems to be better suited than the observation of proton emission to detect 
 the predicted size of the $\gamma$-induced $p\bar{p}$ pair creation process. 
 Facilities that could provide sufficient photon flux 
and are equipped with $4\pi$ detectors for pion detection would 
be CEBAF with the GLAS detector, GRAAL with the BGO detector and ELSA with the SAPHIR 
detector \cite{Grab}. An experimental observation of the $p\bar{p}$ pair creation with 
a reduction of the threshold from 1880 MeV$\approx2M$
to values around 1000 MeV would give strong support to the mean field picture of QHD.

\section*{Acknowledgements}

The authors would like to thank Prof. Robert Vinh Mau for fruitful discussions. 
We also are grateful to Martin Effenberger who helped us with his BUU calculation results. 
One of us (R.H.) wants to acknowledge the support given by the PROCOPE cooperation project (312/pro-u-gg) during  
a research stay in Paris.

\newpage
\begin{appendix}

\section{Calculation of the S-matrix element}

In the following we indicate details for the derivation 
of the S-matrix element of $\gamma$-induced $p\bar{p}$ pair creation. 
The $p$/$\bar{p}$ wave functions appearing in Eq.(\ref{smatkorr}) are given by
\eab
^{(+)}\psi^\mu_p(x)&=&\sqrt{\frac{M}{E({\vec p})V}}\ u({\vec p,\mu})\ e^{-ipx},
\ \ \ \ E(\vec p):=p_0=\sqrt{\vec p^{\ 2}+M^2}\nonumber\\ 
^{(-)}\psi^\rho_{p^\prime}(x)&=&\exp(-iE^\prime x^0) \left(
\begin{array}{ll}
g(r,E^\prime)\ \omega^\rho_\kappa(\theta,\phi)\\
if(r,E^\prime)\ \omega^\rho_{-\kappa}(\theta,\phi)
\end{array}
\right)\ ,\ \ \ \ E^\prime<0\ ,
\eae
where 
\eqb
\omega^\rho_\kappa(\theta,\phi)=\left\{\begin{array}{ll}
\left(\begin{array}{ll}
\frac{\sqrt{j+1-\rho}}{2(j+1)}\ Y^{\rho-\frac{1}{2}}_{j+\frac{1}{2}}(\theta,\phi)\\
-\frac{\sqrt{j+1+\rho}}{2(j+1)}\ Y^{\rho+\frac{1}{2}}_{j+\frac{1}{2}}(\theta,\phi)
\end{array}\right)\ ,\ \ \ \ \ \ \ \kappa>0\\ 
\left(\begin{array}{ll}
\frac{\sqrt{j+\rho}}{2j}\ Y^{\rho-\frac{1}{2}}_{j-\frac{1}{2}}(\theta,\phi)\\
\frac{\sqrt{j-\rho}}{2j}\ Y^{\rho+\frac{1}{2}}_{j-\frac{1}{2}}(\theta,\phi)
\end{array}\right)\ ,\ \ \ \ \ \ \ \ \ \ \ \ \kappa<0\ \ ,
\end{array}\right.
\eqe
\eqb
g(r,E^\prime)=\left\{\begin{array}{ll}
a_1j_{l_\kappa}(cr)\ ,\ \ \ \ \ \ \ \ \ \ \ \ \ \ \ \ \ \  \ 0\le r\le r_0\ ,\\
b_1\sqrt{\frac{2Cr}{\pi}}K_{l_{\kappa}+\frac{1}{2}}(Cr)
\ ,\ \ \ \ \ \ \ r_0<r\ \ ,
\end{array}\right. 
\eqe
\eqb
f(r,E^\prime)=\left\{\begin{array}{ll}
\ \ a_1\frac{\kappa}{|\kappa|}\ \frac{c}{\lambda_1}\ j_{l_{-\kappa}}(cr)
\ ,\ \ \ \ \ \ \ \ \ \ \ \ \ \ \ \ \ 0\le r\le r_0\ ,\\
-b_1\frac{C}{\lambda_2}\sqrt{\frac{2Cr}{\pi}}\ K_{l_{-\kappa}+\frac{1}{2}}(Cr)
\ ,\ \ \ \ \ \ \ \ \ r_0<r\ \ ,
\end{array}\right.
\eqe
\eab
\lambda_1:&=&E^\prime-V+(M-S),\ \ \ \ \ \ \ \ \ \ \lambda_2:=E^\prime+M\ ,\nonumber\\ 
c:&=&\sqrt{(E^\prime-V)^2-(M-S)^2},\ \ C:=\sqrt{{M^2-E^\prime}^2}\ ,\nonumber\\ 
l_\delta:&=&\left\{\begin{array}{ll}
\ \ \delta\ ,\ \ \ \ \ \ \ \ \ \ \ \ \delta>0\ ,\\
-\delta-1\ ,\ \ \ \ \ \ \ \delta<0\ \ .
\end{array}\right.
\eae
The Dirac quantum number $\kappa$ takes the values
\dmb
\kappa=\pm1, \pm2, \pm3,\dots\ ,
\dme
whereas $j$ describes the total angular momentum of the $\bar{p}$ state; 
the functions $j_{n}$ and $K_{n+\frac{1}{2}}$ are the modified spherical Bessel functions. 
The normalization constants $a_1,\ b_1$ are (up to a phase) determined from
\eqb
\int_V d^3x\ \left[^{(-)}\psi^\rho_{p^\prime}
(\vec x)\right]^\dagger\ ^{(-)}\psi^\rho_{p^\prime}(\vec x)=1\ .
\eqe
For the evaluation of Eq.(\ref{smatkorr}) we write
\eab
\label{Scor}
^{cor}S^{1,2;\mu;\rho}_{p,p^\prime}&=&\frac{ie}{V}\ [2\pi\ \delta(E+{\bar E}-k)]\ 
\sqrt{\frac{M}{2k\ E}}\ \bar{u}(\vec p,\mu)\ \ep_\nu^{1,2} 
\left[\gamma^\nu+i\frac{\mu}{2M}k^\prime_\lambda\sigma^{\nu\lambda}\right]\times\nonumber\\ 
& &{4\pi}\ \sum_{l,m}i^l\ {Y^*}^m_l(\hat{\tilde k})\ \int d^3x\ j_l({\tilde k}r)
\ Y^m_l(\theta,\phi)\left(\begin{array}{ll}
g(r,E^\prime)\ \omega^\rho_\kappa(\theta,\phi)\\
if(r,E^\prime)\ \omega^\rho_{-\kappa}(\theta,\phi)
\end{array}
\right)\ \nonumber\\ 
\eae
with
\dmb
\bar E:=|E^\prime|\ ,\ \ \ \vec{\tilde{k}}:=\vec k^\prime-\vec p\ ,
\dme
where time integration has been carried out 
and the exponential $e^{i\vec {\tilde k}\cdot\vec x}$ is expanded in partial waves. With  
\dmb
{Y^*}^m_l(\theta,\phi)=(-1)^m\ {Y}^{-m}_l(\theta,\phi)
\dme
and 
\dmb
\int d\Omega\ {Y^*}^m_l(\theta,\phi)\ Y^{m^\prime}_{l^\prime}(\theta,\phi)=
\delta_{ll^\prime}\ \delta_{mm^\prime}
\dme
the sum in Eq.(\ref{Scor}) degenerates to a single contribution and
 the integral spinor $I(k,\bar E, E, \rho)$ defined by
\eqb
I(k,\bar E, E, \rho):={4\pi}\ \sum_{l,m}i^l\ {Y^*}^m_l(\hat{\tilde k})\ \int d^3x\ j_l({\tilde k}r)
\ Y^m_l(\theta,\phi)\left(\begin{array}{ll}
g(r,E^\prime)\ \omega^\rho_\kappa(\theta,\phi)\\
if(r,E^\prime)\ \omega^\rho_{-\kappa}(\theta,\phi)
\end{array}
\right)
\eqe
is given for $\kappa>0$ as
\eab
\label{ints}
I(k,\bar E, E, \rho)&=&4\pi\{ a_1\  \int_0^{r_0} dr\ r^2\times\nonumber\\ 
& &\left(\begin{array}{llll}
i^{j+\frac{1}{2}}\ (-1)^{-(\rho-\frac{1}{2})}\ {Y^*}^{-(\rho-\frac{1}{2})}_{j+\frac{1}{2}}(\hat{\tilde k})\ 
\frac{\sqrt{j+1-\rho}}{2(j+1)}\ j_{j+\frac{1}{2}}({\tilde k}r)\  j_{l_\kappa}(cr)\\ 
i^{j+\frac{1}{2}}\ (-1)^{-(\rho+\frac{1}{2})+1}\ {Y^*}^{-(\rho+\frac{1}{2})}_{j+\frac{1}{2}}(\hat{\tilde k})\ 
\frac{\sqrt{j+1+\rho}}{2(j+1)}\ j_{j+\frac{1}{2}}({\tilde k}r)\  j_{l_\kappa}(cr)\\ 
i^{j-\frac{1}{2}+1}\ \frac{\kappa}{|\kappa|}\  \frac{c}{\lambda_1}\ 
(-1)^{-(\rho-\frac{1}{2})}\ {Y^*}^{-(\rho-\frac{1}{2})}_{j-\frac{1}{2}}(\hat{\tilde k})\ 
\frac{\sqrt{j+\rho}}{2j}\ j_{j-\frac{1}{2}}({\tilde k}r)\  j_{l_{-\kappa}}(cr)\\ 
i^{j-\frac{1}{2}+1}\ \frac{\kappa}{|\kappa|}\  \frac{c}{\lambda_1}\ 
(-1)^{-(\rho+\frac{1}{2})}\ {Y^*}^{-(\rho+\frac{1}{2})}_{j-\frac{1}{2}}(\hat{\tilde k})\ 
\frac{\sqrt{j-\rho}}{2j}\ j_{j-\frac{1}{2}}({\tilde k}r)\  j_{l_{-\kappa}}(cr)
\end{array}
\right)+\nonumber\\ 
& &b_1 \int_{r_0}^{\infty} dr\ r^2 \sqrt{\frac {2Cr}{\pi}}\times\nonumber\\ 
& &\left(\begin{array}{llll}
i^{j+\frac{1}{2}}\ (-1)^{-(\rho-\frac{1}{2})}\ {Y^*}^{-(\rho-\frac{1}{2})}_{j+\frac{1}{2}}(\hat{\tilde k})\ 
\frac{\sqrt{j+1-\rho}}{2(j+1)}\ j_{j+\frac{1}{2}}({\tilde k}r)\  K_{l_\kappa+\frac{1}{2}}(Cr)\\ 
i^{j+\frac{1}{2}}\ (-1)^{-(\rho+\frac{1}{2})+1}\ {Y^*}^{-(\rho+\frac{1}{2})}_{j+\frac{1}{2}}(\hat{\tilde k})\ 
\frac{\sqrt{j+1+\rho}}{2(j+1)}\ j_{j+\frac{1}{2}}({\tilde k}r)\  K_{l_\kappa+\frac{1}{2}}(Cr)\\ 
i^{j-\frac{1}{2}+1}\ \frac{C}{\lambda_2}\ (-1)^{-(\rho-\frac{1}{2})+1}\ {Y^*}^{-(\rho-\frac{1}{2})}_{j-\frac{1}{2}}(\hat{\tilde k})\ 
\frac{\sqrt{j+\rho}}{2j}\ j_{j-\frac{1}{2}}({\tilde k}r)\  K_{l_{-\kappa+\frac{1}{2}}}(Cr)\\ 
i^{j-\frac{1}{2}+1}\ \frac{C}{\lambda_2}\ (-1)^{-(\rho+\frac{1}{2})+1}\ {Y^*}^{-(\rho+\frac{1}{2})}_{j-\frac{1}{2}}(\hat{\tilde k})\ 
\frac{\sqrt{j-\rho}}{2j}\ j_{j-\frac{1}{2}}({\tilde k}r)\  K_{l_{-\kappa+\frac{1}{2}}}(Cr)
\end{array}
\right)\}_{\ \ \ .}
\nonumber\\ \nonumber\\ 
\eae
A similar expression results for the case $\kappa<0$. 
The integrals in Eq.(\ref{ints}) are evaluated numerically.

Finally, the bilinear expression 
\eqb
\label{bilin}
B(k,\bar E, E, \mu, \rho):=\bar{u}(\vec p,\mu)\ \ep_\nu^{1,2}\ 
\left[\gamma^\nu+i\frac{\mu_{a.m.}}{2M}k_\lambda\sigma^{\nu\lambda}\right]\ I(k,\bar E, E, \rho)
\eqe
is easily calculated.

\section{Rosenbluth's formula}

Here we show that the influence of static external potentials ($V_\mu, \phi$) 
does not alter the form of the free corrected transition current 
\eqb
\label{rbf}
j_{tr, cor}^{\mu}=\bar{\psi_{p^\prime}}\gamma^{\mu}\psi_p +\frac{\mu_{a.m.}}{2M}
\pd_\nu\bar{\psi_{p^\prime}}\sigma^{\mu\nu}\psi_p \ ,
\eqe
where $\psi_{p^\prime}$ and $\psi_p$ are solutions of the free Dirac equation. 
 
If $\psi_{p^\prime}$ and $\psi_p$ are solutions of the Dirac equation with static scalar and vector potentials, the corrected
transition current is required to fulfill the following demands:\\ \\ 
 {\bf{1}}) the corrected current has to transform like a Lorentz {\em vector} 
  under the whole Lorentz group and has to be a 
  bilinear form in the fields $\psi_{p^\prime}$ and $\psi_p$, \\  
 {\bf{2}}) it has to be conserved \\  
 {\bf{3}}) and asymptotically, that is $V_\mu, \phi \longrightarrow 0$, given by Eq.(\ref{rbf}).\\ 

In addition to the bilinear covariants of the Dirac theory we have the four gradient, 
the potentials themselves and the mass to 
construct a corrected current. The most general ansatz consistent with {\bf{1}}) is then
\eqb
\label{stans}
j^\mu_{tr, cor}=^S F^\mu \bar{\psi_{p^\prime}}\psi_p+
^V F\bar{\psi_{p^\prime}}\gamma^\mu\psi_p+
^T F_\nu\bar{\psi_{p^\prime}}\sigma^{\mu\nu}\psi_p
\eqe
where the expressions $^S F^\mu$, $^V F$, $^T F_\nu$ depend on the four 
Lorentz covariants mentioned only. In accordance with demand {\bf{1}}) 
one can split up for instance $^S F^\mu$ in the following way
\eab
\label{split}
^S F^\mu&=&^S F^\mu(U^\kappa, \pd^\lambda, \sigma, M)\nonumber\\
&=&\left\{^S F^{D}\frac{1}{M}\pd^\mu+^S F^{D^2D}\frac{1}{M^3}\pd_\kappa\pd^\kappa \pd^\mu+
{\cal O}(\left(\frac{1}{M}\right)^5)\right\}+\nonumber\\
& &\left\{\frac{1}{M^2}[^S F^{\sigma D}\sigma\pd^\mu+^S F^{D\sigma}(\pd^\mu\sigma)]+
{\cal O}(\left(\frac{1}{M}\right)^3)\right\}
+\nonumber\\ 
& &\left\{\frac{1}{M}\ ^S F^{V}V^\mu+\frac{1}{M^2}\ ^S F^{\sigma V}\sigma V^\mu+
\frac{1}{M^3}[^S F^{V^2V}V_\kappa V^\kappa V^\mu+\right.\nonumber\\ 
& &\left.^S F^{\sigma^2 V}\sigma^2 V^\mu]+
{\cal O}(\left(\frac{1}{M}\right)^4)\right\}
\eae
with similar expressions for $^V F$ and $^T F_\nu$. The $F^{DD^i}, F^{\dots P^jD^i\dots}$,
 $F^{P^i}$ ($P=\sigma, V$) denote dimensionless scalar factors. 
 In Eq.(\ref{split}) superscripts $DD^i,\ \dots P^jD^i\dots\ , P^i$ 
indicate that in the corresponding curly brackets an expansion with respect to 
derivatives, derivatives and potentials, potentials only is performed. In the decomposition of 
$^V F$ there also appears a "constant" term $^V F^0$.  

For the case of $^S F^\mu$ and $^V F$ the $D$-terms violate demand {\bf{3}}) and therefore cannot
 contribute as well as all the $D$-terms in the decomposition of $^T F_\nu$ except for $^T F^{D^1}\pd_\nu$. 

If we take the divergence of Eq.(\ref{stans}) the terms with $^V F^0$ and 
$^T F^{D^1}\pd_\nu$ vanish by virtue of the equation of motion and the antisymmetry of $\sigma^{\mu\nu}$.
 The imposition of conditions {\bf{2}}) and {\bf{3}}) leads to  
\eab
\label{group}
0=\pd_\mu j_{tr, cor}^\mu&=&\bar{\psi_{p^\prime}}\left 
({\leftrightarrow\atop{\pd_\mu\  ^S \tilde{F}^\mu}}+{\leftrightarrow\atop{\pd_\mu\  ^V \tilde{F}}\gamma^\mu}+
{\leftrightarrow\atop{\pd_\mu\  ^T \tilde{F}_\nu}\sigma^{\mu\nu}} \right )\psi_p + \nonumber\\ 
& &\pd_\mu\bar{\psi_{p^\prime}}\left 
({\leftrightarrow\atop{^S \tilde{F}}^\mu}+{\leftrightarrow\atop{^V \tilde{F}}\gamma^\mu}+
{\leftrightarrow\atop{^T \tilde{F}}_\nu\sigma^{\mu\nu}} \right )\psi_p+ \nonumber\\ 
& &\bar{\psi_{p^\prime}}\left 
({\leftrightarrow\atop{^S \tilde{F}}^\mu}+{\leftrightarrow\atop{^V \tilde{F}}\gamma^\mu}+
{\leftrightarrow\atop{^T \tilde{F}}_\nu\sigma^{\mu\nu}} \right )\pd_\mu\psi_p\ ,
\eae
where the symbol  " $\leftrightarrow\atop$ "  stands for the right arrangement of the
unsaturated derivatives with respect to the product rule , for example
\eqb
{\leftrightarrow\atop{\pd_\kappa\pd^\kappa}}
:={\leftarrow\atop{\pd_\kappa}} {\leftarrow\atop{\pd^\kappa}}+
{\leftarrow\atop{2\ \pd_\kappa}} {\rightarrow\atop\pd^\kappa}+
{\rightarrow\atop{\pd_\kappa}} {\rightarrow\atop{\pd^\kappa}}\ ,
\eqe
and " $\tilde{}$ " indicates that we already have omitted the terms forbidden by {\bf 3}) and those with $^V F^0$ and 
$^T F^{D^1}\pd_\nu$ ($\nu=0,\dots,3$).

Each term in the sums over $\mu$ in Eq.(\ref{group}) must vanish by itself;
 since $\bar{\psi_{p^\prime}}$ and $\psi_p$ are
arbitrary states of the spectrum, the corresponding sum of the bilinears in the brackets must vanish. 
But because of the linear independence of the Dirac matrices ${\bf 1}, \gamma^\mu, \sigma^{\mu\nu}$ 
\eqb
\pd_\mu ^S \tilde{F}^\mu=\pd_\mu ^V \tilde{F}\gamma^\mu=
\pd_\mu ^T \tilde{F}_\nu\sigma^{\mu\nu}
=^S \tilde{F}^\mu=^V \tilde{F}=^T \tilde{F}_\nu=0\ \ \ \ \forall \mu, \nu \ \ \mbox{(no sum over $\mu$ and $\nu$)}\ .
\eqe
 
To summarize we have found that in the presence of static 
external scalar and vector potentials Rosenbluth's formula for the transition current still holds.

\section{Cross sections} 
 
With the calculated S-matrix element of Eq.(\ref{Scor}) the corresponding 
differential cross section $\frac{d\sigma}{d\Omega}$ can be easily obtained. Here we average 
over the two photon polarizations $1,2$ and sum over the angular 
momentum projections of the $p$ and $\bar{p}$ states. With phase space $V\frac{d^3p}{(2\pi)^3}$ and 1 for 
the continuum $p$ and bound $\bar{p}$ state, respectively, the differential cross section is given as
\eab 
\label{cs}
\frac{d\sigma}{d\Omega}&=&\sum_{1,2;\mu;\rho}\frac{1}{2}\frac{e^2}{V^2}\ 2\pi\ \delta(E+{\bar E}-k)\ 
\frac{M}{2k E}|B(k,\bar E, E, \mu, \rho)|^2\ V^2\frac{p^2\ dp}{(2\pi)^3}\nonumber\\ 
&=&\sum_{1,2;\mu;\rho}\frac{1}{4}\frac{4\pi\alpha}{k E}\ 2\pi\ \delta(E+{\bar E}-k)\ 
M\ |B(k,\bar E, E, \mu, \rho)|^2\ \frac{p^2\ dp}{(2\pi)^3}\ \nonumber\\ 
&=&\sum_{1,2;\mu;\rho}\frac{1}{2}\frac{\alpha}{k E}\ \delta(E+{\bar E}-k)\ 
M\ |B(k,\bar E, E, \mu, \rho)|^2\ \frac{p\ E\ dE}{(2\pi)}\ \nonumber\\
&=&\sum_{1,2;\mu;\rho}\frac{1}{4}\frac{\alpha}{\pi k}\ M\ |B(k,\bar E, E, \mu, \rho)|^2\ p|_{E=k-\bar E}\ \nonumber\\
&=&\sum_{1,2;\mu;\rho}\frac{1}{4}\frac{\alpha}{\pi k}\ M\ |B(k,\bar E, \mu, \rho)|^2\ \sqrt{(k-\bar E)^2-M^2}\ , 
\eae
where
\dmb
\alpha\approx\frac{1}{137}\ .
\dme
The total cross section $\sigma$ is obtained by (numerically) integrating Eq.(\ref{cs}) over the angle 
$\theta$ and multiplying with $2\pi$. 

Inclusive cross sections are obtained by summing over all 
possible final state contributions at a given $\gamma$-energy $k$ with 
\eqb
\label{ics}
\sigma_{inc}(k)=\sum_{{\bar E\atop 
\bar E+E=k}\atop 
E\ge M }
\sigma(\bar E,k)\ \ .
\eqe
 
\end{appendix}

\newpage
\noindent
Figure 1:\ {Schematic view of photon induced $p\bar{p}$ pair creation 
in the presence of a scalar-vector mean field potential.}
\\ 
\\ 
Figure 2:\ {Negative energy spectrum of the Dirac equation Eq.(\ref{barem}) with 
the mean field potential parameters 
$S=400$ MeV, $V=450$ MeV and the nuclear radius $r_0=7.2$ fm corresponding to $^{208}$Pb.}
\\ 
\\
Figure 3:\ {Dependence of the cross section $\sigma$ for direct $p$ emission 
due to $\gamma$-induced $p\bar{p}$ pair creation on the potential depth $S=V+50$ MeV in steps of 10 MeV.
 The energy 
of the proton is fixed at $M+8$ MeV (M is the nucleon mass) while the negative energy solution 
is in the highest $\kappa=+1$ state corresponding to a $\bar{p}$ 
in the lowest lying $\kappa=-1$ state. 
The radius $r_0=7.2$ fm adjusted to 
$^{208}$Pb is kept fixed.}
\\ 
\\
Figure 4:\ {Differential inclusive cross section $\frac{d\sigma_{inc}}{d\Omega}$ for direct $p$ emission due 
to $p\bar{p}$ pair creation at a photon energy of $k=k_{th}+50$ MeV  (dashed) and of 
$k=k_{th}+30$ MeV (solid) with the threshold energy $k_{th}=1035$ MeV. Nuclear potential parameters are 
$S=450$ MeV, $V=400$ MeV and $r_0=7.2$ fm for $^{208}$Pb.}
\\ 
\\
Figure 5:\ {Inclusive $p\bar{p}$ pair creation cross section $\sigma_{inc}$ including negative energy states
 up to $\kappa=+7$ in dependence on the photon energy $k$. Mean
field potential parameter values are as in Fig. 4. 
The threshold for photoproduction of $p\bar{p}$ pairs lies for these potential
parameters at $k_{th}=1035$ MeV.}
\\ 
\\
\newpage
\begin{figure}[h]
\label{1}
\vskip 15cm 
\includegraphics{fig1.ps}
\caption{}
\end{figure}

\newpage

\begin{figure}[h]
\label{2}
\vskip 15cm 
\includegraphics{fig2.ps}
	 \caption{} 
\end{figure}

\newpage
  
\begin{figure}[h]
\label{3}
\vskip 10.5cm 
\includegraphics{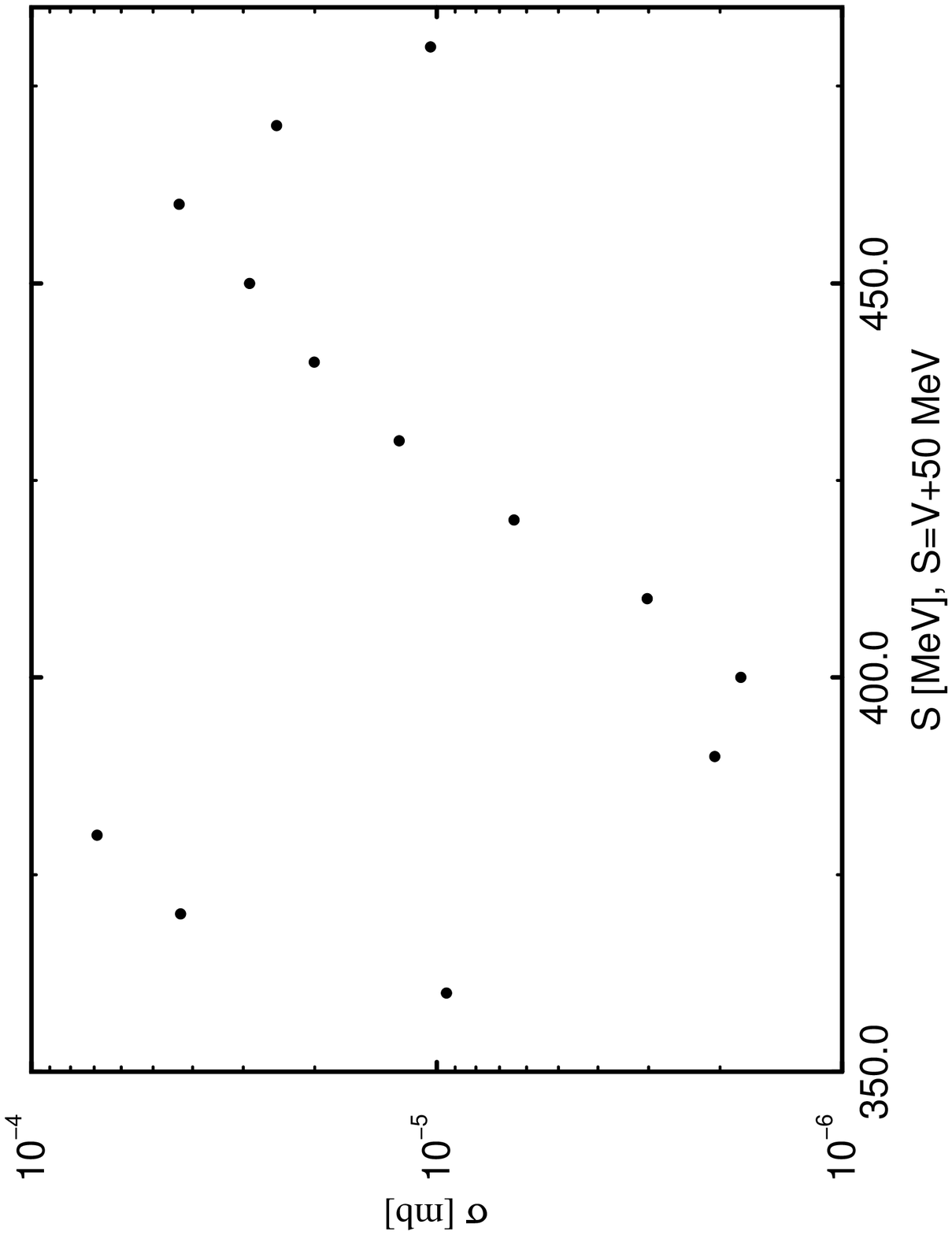}
\caption{} 
\end{figure}

\newpage

\begin{figure}[h]
\label{4}
\vskip 10.5cm 
\includegraphics{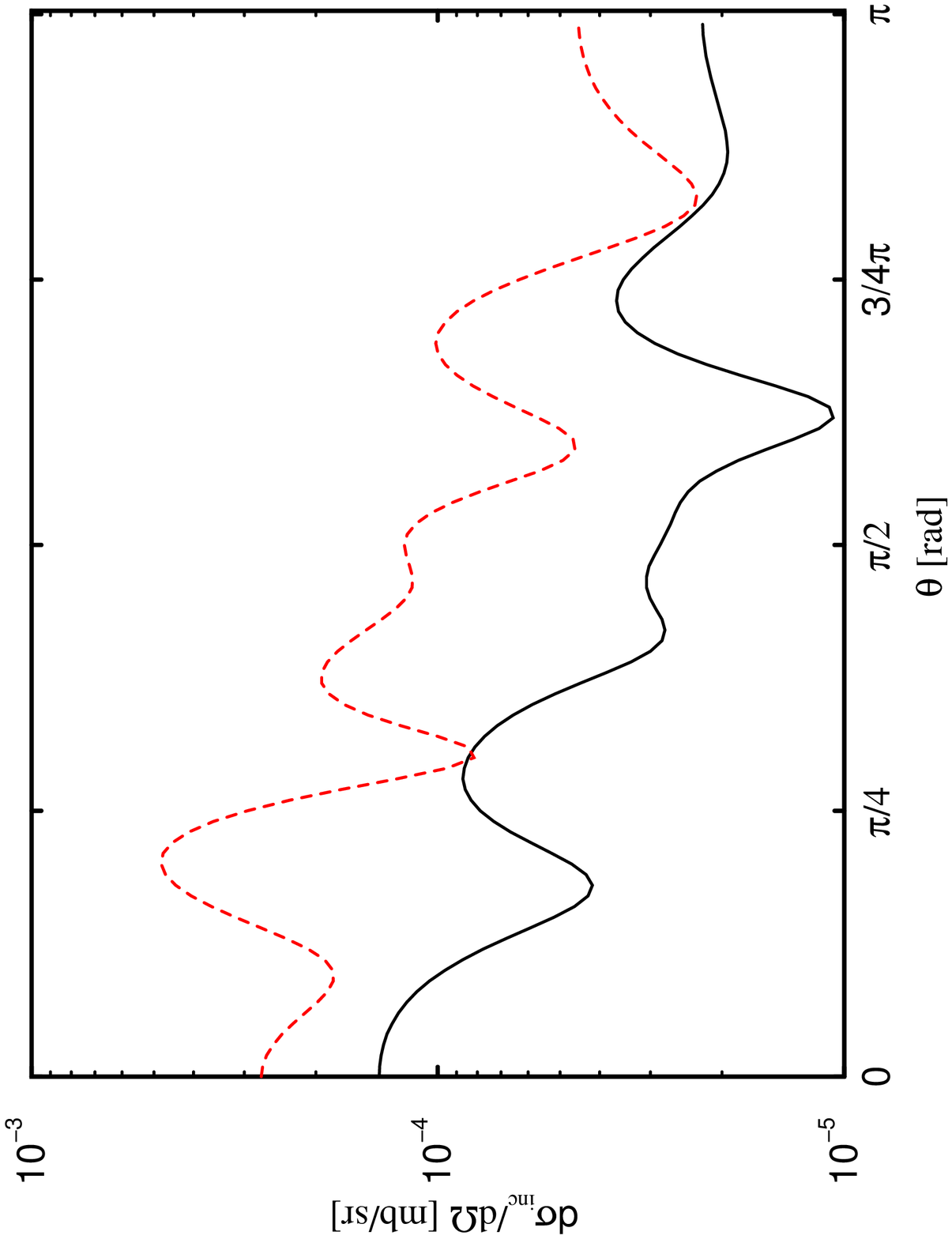}
\caption{} 
\end{figure}

\newpage

\begin{figure}[h]
\label{5}
\vskip 10.5cm 
\includegraphics{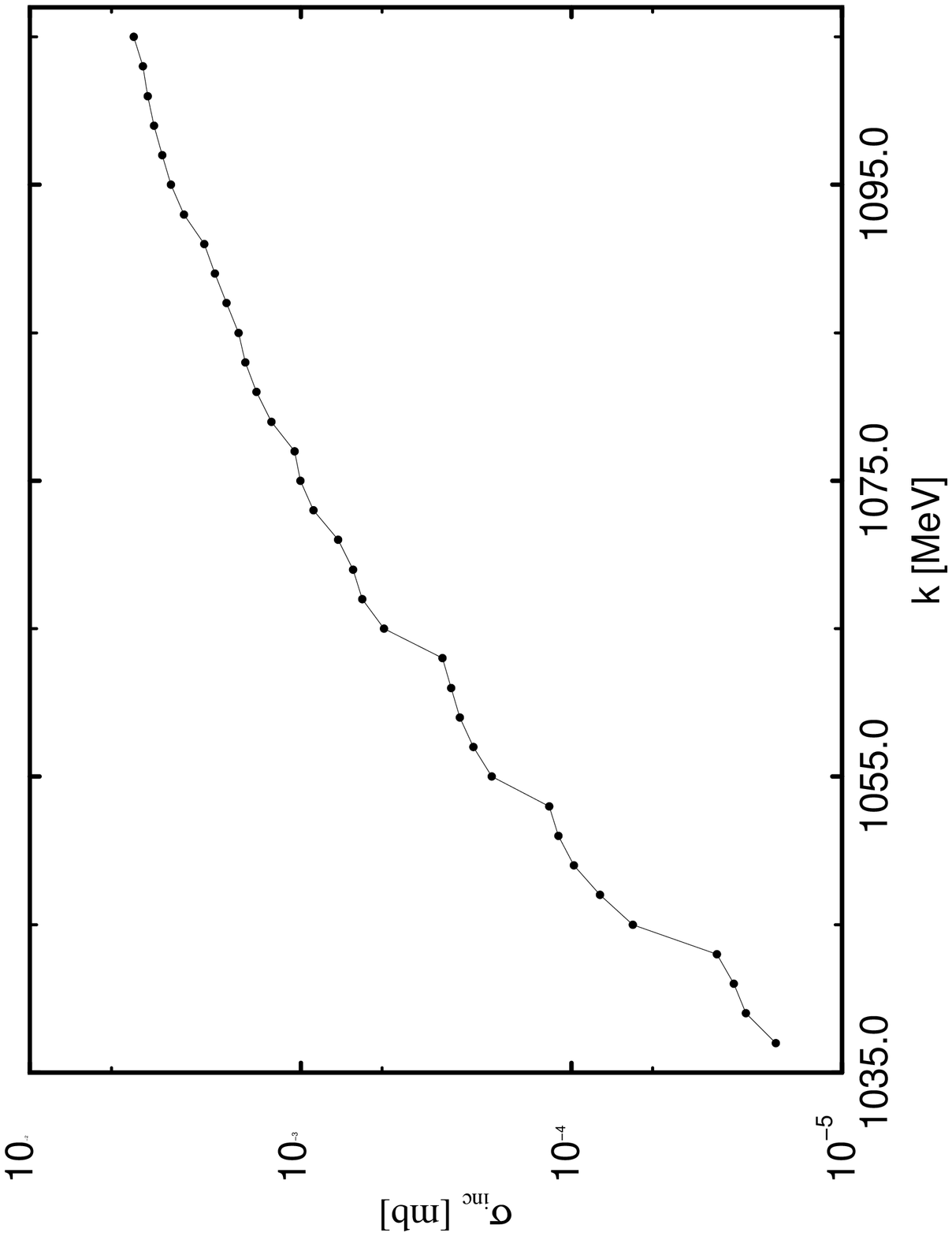}
\caption{} 
\end{figure}

\end{document}